# Density Functional Theory Calculations for Spin Crossover Complexes

HAUKE PAULSEN AND ALFRED X. TRAUTWEIN

*Institut für Physik*
*Universität zu Lübeck*
*Ratzeburger Allee 160*
*D-23538 Lübeck*
*Germany*

e-mail: paulsen@physik.mu-luebeck.de
phone: (+49) (0) 451 500 4208
FAX: (+49) (0) 451 500 4214

Density functional theory (DFT) provides a theoretical framework for efficient and fairly accurate calculations of the electronic structure of molecules and crystals. The main features of density functional theory are described and DFT methods are compared with wavefunction-based methods like the Hartree-Fock approach. Some recent applications of DFT to spin crossover complexes are reviewed, e.g. the calculation of vibrational modes and of differences of entropy, vibrational energy, and total electronic energy.

*Keywords: density functional theory, spin crossover, vibrational modes, transition temperature*

# 1 Introduction

The spin crossover (SCO) phenomenon can be explained qualitatively by ligand-field theory [1]. Together with empirical parameters, the ligand-field model gives also quantitative explanations and even has some predictive power. The most important merit of ligand-field theory is probably, that it provides an easy to understand model. Instead, *ab initio* or *first principles*[1] methods for calculating the electronic structure in a sense resemble a black box: all calculated quantities are automatically derived from the Schrödinger equation. Electronic structure calculations are, therefore, not meant to replace the ligand-field model but to complement it. These methods are most useful, if quantitative accuracy is needed or if quantities that have not been measured are to be predicted.

With respect to electronic-structure calculations, transition-metal complexes are intricate objects. This is especially true for SCO complexes where the almost vanishing energy difference between different spin multiplicities results from a delicate balance of various factors. For exactly this reason these complexes are very interesting test objects for judging the quality of theoretical methods, in particular, if these methods can predict the true ground state multiplicity. Methods based on density functional theory (DFT) are currently the natural choice for calculating the electronic structure of transition-metal complexes. DFT methods are efficient enough to handle large molecules (e.g. containing one hundred atoms as an order of magnitude) and they provide fairly accurate results for many quantities. Unlike DFT methods wavefunction-based methods can, in principle, produce any desired accuracy. However, SCO complexes are by far too large to be treated with sophisticated wavefunction-based methods. Those methods that can actually be applied to SCO complexes, like the Hartree-Fock approach, yield poor results in comparison to DFT methods.

This article is divided in two parts. In the first part (section 2) density functional theory will be introduced and the most important DFT methods and some recent developments that are especially relevant to SCO complexes will be discussed. The second part (section 3) reviews some applications related to SCO compounds, i.e. examples are described for the calculation of molecular geometry, Mössbauer parameters, normal modes of vibrations and IR, Raman, and nuclear inelastic scattering (NIS [2]) intensities. From the normal modes of vibrations, the vibrational energy and entropy differences can be derived with reasonable accuracy. One of the quantities of SCO complexes that is a challenge to calculate is the total electronic energy difference $\Delta E$ between the high-spin (HS) and the low-spin (LS) state; yet, this difference has been calculated for several complexes with at least reasonable accuracy. Finally, calculations for iron(II) SCO complexes with substituted tris(pyrazolyl) ligands are presented; they provide an example of how DFT calculations might be used to support the design of future SCO complexes with specified properties.

---

[1] Since there is no consensus in literature on whether DFT methods should be called *ab initio* or *first principles*, these terms will be avoided in the following.

## 2 Density Functional Theory

The electronic ground state of a system containing $N$ electrons is usually described by the many-electron wavefunction $\Psi(\mathbf{x}_1,\mathbf{x}_2,\ldots,\mathbf{x}_N)$, which is a solution of the many-electron Schrödinger equation (the coordinates $\mathbf{x}_i$ comprise space coordinates $\mathbf{r}_i$ and spin coordinates $s_i=\pm\frac{1}{2}$). The central point of density functional theory (see [3-9] for an overview) is the observation that the ground state can be equivalently described by the one-particle charge density $\rho(\mathbf{r})$, which is defined by

$$\rho(\mathbf{r}_1) = e \sum_{s_1} \int |\Psi(\mathbf{x}_1,\mathbf{x}_2,\ldots,\mathbf{x}_N)|^2 \, d\mathbf{x}_2\ldots d\mathbf{x}_N \, , \qquad (1)$$

where the integration over $\mathbf{x}_i$ includes integration over $\mathbf{r}_i$ and summation over $s_i$. Modern DFT started with the theorems of Hohenberg and Kohn[2] in 1964. The first Hohenberg-Kohn theorem [10] states that there is a one-to-one mapping between the ground-state charge density $\rho_0(\mathbf{r})$, the external potential of the nuclei $v_{\text{ext}}(\mathbf{r})$ and the ground-state wavefunction $\Psi_0$

$$\rho_0(\mathbf{r}) \leftrightarrow v_{\text{ext}}(\mathbf{r}) \leftrightarrow \Psi_0 \, . \qquad (2)$$

While it is intuitively obvious that the external potential determines the wavefunction, which in turn determines the charge density, it was unknown before that for each (non-degenerate) ground-state density $\rho_0(\mathbf{r})$ there exists exactly one ex-ternal potential $v_{ext}(\mathbf{r})$. The first Hohenberg-Kohn theorem is valid for any kind of local external potential. In the special but most important case, that $v_{ext}(\mathbf{r})$ describes the nuclear electrostatic potential of a molecule containing $M$ nuclei with atomic numbers $Z_n$ and positions $\mathbf{R}_n$,

$$v_{ext}(\mathbf{r}) = -\frac{1}{4\pi\varepsilon_0}\sum_{n=1}^{M}\frac{eZ_n}{|\mathbf{R}_n - \mathbf{r}|} \quad , \qquad (3)$$

the theorem can be understood by very intuitive arguments that are attributed [11] to E. B. Wilson: if $\rho_0(\mathbf{r})$ is known, $Z_n$ and $\mathbf{R}_n$ can be derived from the cusps of $\rho_0(\mathbf{r})$; thus $v_{ext}(\mathbf{r})$ is defined and the number of electrons $N$ can be determined by integration of $\rho_0(\mathbf{r})$ over space. The second Hohenberg-Kohn theorem [10] states that for a given external potential $v$ there exists a functional $E_v[\rho]\geq E_0$ that yields minimum energy for the ground state density $\rho_0(\mathbf{r})$. The functional for the total energy can be written as a sum of two functionals with quite different properties: $E_v[\rho]=E_{HK}[\rho]+E_{ne}[\rho]$. The first term, $E_{ne}[\rho]=\int v_{ext}(\mathbf{r})\rho(\mathbf{r})d\mathbf{r}$, describes the interaction of the electrons with the external potential and depends on the particular molecule, but it is straightforward to calculate. The second term, the Hohenberg-Kohn functional $E_{HK}[\rho]$, is a universal functional, i.e. it does not depend on the particular system. Unfortunately, only the existence of this functional can be proven, but an exact expression for $E_{HK}[\rho]$ is not known. First attempts to realize such a functional in an approximate way were made already in the early stage of quantum mechanics by Thomas and Fermi (see Ref. [4] for a review), long before density functional theory was founded. They derived an expression for the electronic energy $E_{TF}[\rho]=\int[t(\mathbf{r})+v_{ext}(\mathbf{r})+j(\mathbf{r})]\rho(\mathbf{r})d\mathbf{r}$ that uses local potentials: $t(\mathbf{r})=(3h^2/10me)(3\rho(\mathbf{r})/8\pi e)^{2/3}$ for the kinetic energy, $v_{ext}(\mathbf{r})$ for the external potential, and

$$j(\mathbf{r}) = \frac{1}{4\pi\varepsilon_0}\int\frac{\rho(\mathbf{r}')}{|\mathbf{r}'-\mathbf{r}|}d\mathbf{r}' \qquad (4)$$

for the classical Coulomb self-interaction of a given charge density. In quantum mechanics $j(\mathbf{r})$ describes which effective potential is seen by a reference electron, that moves in the average Coulomb field of all electrons (including the reference electron, thus including unwanted self-interaction of the reference electron). The Thomas-Fermi model is the first example of a density functional method, but due to the rather coarse approximation of the kinetic energy, this model never gained importance for applications to real systems. Surprisingly, the way to improve the kinetic-energy functionals leads back to wavefunctions. For this reason the Hartree-Fock method, the fundamental wavefunction based procedure for electronic-structure calculations, will be reviewed in the next section.

---

[2] Walter Kohn was awarded the Nobel Prize for chemistry in 1998 for his contributions to density functional theory.

## 2.1 Hartree-Fock Method

One of the earliest attempts in quantum chemistry has been made by Hartree [12] writing the many-electron wavefunction of an ion as the product of single-particle functions, called orbitals, $\Psi(\mathbf{x}_1,\mathbf{x}_2,\ldots,\mathbf{x}_N)=\psi_1(\mathbf{x}_1)\psi_2(\mathbf{x}_2)\ldots\psi_N(\mathbf{x}_N)$. If $\Psi$ is written in this form, known as the Hartree product, the probability of finding a particular electron $i$ at position $\mathbf{r}_i$ does not depend on the positions or the spins of the other electrons, in other words the electron positions and spins are not correlated. Using the effective potential $j(\mathbf{r})$ for the electron-electron interaction, a set of one-electron Schrödinger equations

$$\left[-\frac{\hbar^2}{2m}\nabla^2 + ev(\mathbf{r}) + ej(\mathbf{r}) - \hat{k}(\mathbf{x})\right]\psi_i(\mathbf{x}) = \varepsilon_i\psi_i(\mathbf{x}) \; , \qquad (5)$$

can be formulated, called canonical Hartree-Fock equations (see e.g. [4,13]), where the first operator in parentheses stands for the kinetic energy. The operator $\hat{k}(\mathbf{x})$, which is defined by

$$\hat{k}(\mathbf{x})\psi_i(\mathbf{x}') = \frac{e^2}{4\pi\varepsilon_0}\sum_{k=1}^{N}\int\frac{\psi_k^*(\mathbf{x}')\psi_k(\mathbf{x})\psi_i(\mathbf{x}')}{|\mathbf{r}'-\mathbf{r}|}d\mathbf{x}' \; , \qquad (6)$$

has to be included, if the antisymmetry of the wavefunction has to be considered, that means $\Psi(\ldots,\mathbf{x}_i,\ldots,\mathbf{x}_k,\ldots) = -\Psi(\ldots,\mathbf{x}_k,\ldots,\mathbf{x}_i,\ldots)$ for arbitrary electrons $i$ and $k$. This can be reached by forming a suitable linear combination of the Hartree product and its permutations with respect to the electron coordinate, e.g. $\Psi(\mathbf{x}_1,\mathbf{x}_2) = [\psi_1(\mathbf{x}_1)\psi_2(\mathbf{x}_2)-\psi_1(\mathbf{x}_2)\psi_2(\mathbf{x}_1)]/\sqrt{2}$, with the generalization to the $N$-electron case, $\Psi=\det[\psi_i(\mathbf{x}_k)]/\sqrt{N!}$, known as the Slater determinant. Since $j(\mathbf{r})$ depends on the orbitals of all other electrons $k\neq i$, Eq. (5) has to be solved iteratively until a self-consistent set of orbitals $\psi_i$, $i=1\ldots N$, has been reached. It can be shown that this procedure leads to the lowest total energy $E$ that is possible for a trial function in the form of a Slater determinant or in the form of a Hartree product, depending on whether the operator $\hat{k}(\mathbf{x})$ is included in the one-electron Schrödinger equation or not. In the first case, the procedure is the well-known Hartree-Fock method; in the latter case, it is the Hartree approximation. The total electronic energy $E$ is not equal to the sum of the one-electron energies $\varepsilon_i$, since otherwise the electron-electron interaction between the electrons would be counted twice. The total energy can be written as a sum of one kinetic-energy term and of three potential-energy terms $E=E_{kin}+E_{ne}+E_{Coul}-E_x$, with

$$E_{kin} = -\frac{\hbar^2}{2m} \sum_{i=1}^{N} \int \psi_i^*(\mathbf{r}) \nabla^2 \psi_i(\mathbf{r}) d\mathbf{r} \qquad (7)$$

$$E_{ne} = \int v(\mathbf{r}) \rho(\mathbf{r}) d\mathbf{r} , \qquad (8)$$

$$E_{Coul} = \frac{1}{4\pi\varepsilon_0} \iint \frac{\rho(\mathbf{r'})\rho(\mathbf{r})}{|\mathbf{r'}-\mathbf{r}|} d\mathbf{r'} d\mathbf{r} , \qquad (9)$$

$$E_x = \frac{e^2}{4\pi\varepsilon_0} \sum_{i,k}^{N} \iint \frac{\psi_i^*(\mathbf{r})\psi_k^*(\mathbf{r'})\psi_i(\mathbf{r'})\psi_k(\mathbf{r})}{|\mathbf{r'}-\mathbf{r}|} d\mathbf{r'} d\mathbf{r} . \qquad (10)$$

$E_{ne}$ and $E_{Coul}$ represent the classical potential energy of a charge distribution $\rho(\mathbf{r})$ in an external potential $v_{ext}(\mathbf{r})$. The third contribution to the potential energy, the exchange energy $E_x$, has no analogue in classical physics and is not present, if the wavefunction is written as a Hartree product. In the Hartree approximation the antisymmetry of the wavefunction is taken into account only by the Pauli principle, not allowing two electrons of the same spin to have the same orbital. That means a particular electron is assigned to a particular orbital, contrary to the fact that electrons are indistinguishable. As for the example of the He $^1S$ ground state and the corresponding $^3S$ excited state it becomes clear that the total electronic energies calculated in the Hartree approximation[3] (-5.105 and -3.057 MJ mol$^{-1}$, respectively) are only a first, qualitative approximation to the experimental values (-7.624 and -5.721 MJ mol$^{-1}$ [15]). One reason for the failure of quantitative agreement is the presence of self-interaction of each electron in the effective potential $j(\mathbf{r})$. The failure for the $^3S$ excited state is also due to the fact that the Hartree product is not antisymmetric.

---

[3] This work, using the cc-VQZ basis set [14]

In the Hartree-Fock method, the self-interaction of the individual electrons cancels, since it is included in the potential $j(\mathbf{r})$ and in the operator $\hat{k}(\mathbf{x})$. This cancellation of self-interaction is a very fortunate property of the Hartree-Fock method. For DFT methods, the incomplete cancellation of self-interaction is a problem, because the classical Coulomb interaction $E_{\text{Coul}}$ is calculated exactly, whereas the exchange energy $E_x$ is calculated only approximately. The inclusion of exchange leads to calculated energies for the He $^1S$ ground state and $^3S$ excited state (-7.513 and -5.229 MJ mol$^{-1}$, respectively) in reasonable agreement with the experimental values. The remaining difference between the calculated and the measured energies is mainly due to the electron correlation that is missing, because the wavefunction is written as a Hartree product or as a Slater determinant. Taking additionally into consideration the correlation of the positions of electrons leads to a reduction of the total electronic energy since the electrons can avoid each other. The correlation energy can be defined as difference between the true energy $E$ and the Hartree-Fock energy $E_{\text{HF}}$

$$E_c = E - E_{\text{HF}} \ . \qquad (11)$$

Taking into account the Coulomb correlation, the monodeterminantal wavefunction must be replaced by a linear combination of Slater determinants. Using such an approach the energy of the He $^1S$ ground state can be lowered by 107 kJ mol$^{-1}$, reaching thus almost the experimental value. The remaining difference of 4 kJ mol$^{-1}$ is mostly due to relativistic effects and the finite mass of the He nucleus.

## 2.2 Local Density Approximation

The first density functional method that successfully describes real systems was introduced by Slater [16]. His idea was to replace the computational demanding Hartree-Fock exchange by the approximation $E_x[\rho]=\int v_x(\mathbf{r})\rho(\mathbf{r})d\mathbf{r}$ with

$$v_x(\mathbf{r}) = -\frac{9\alpha}{32\pi\varepsilon_0}\left[\frac{3e^2}{\pi}\rho(\mathbf{r})\right]^{1/3} . \qquad (12)$$

This approach has been called $X\alpha$ method after the semiempirical parameter $\alpha$, which has been introduced later by Schwartz [17]. Although, originally meant to approximate the Hartree-Fock approach, the $X\alpha$ method turned out to be superior in some areas like solid-state physics or transition-metal chemistry. Kohn and Sham established a sound theoretical basis for this method later: they proposed a system of non-interacting electrons in order to calculate accurately the kinetic energy in the DFT framework [18]. These non-interacting electrons move in the classical Coulomb potential of the nuclei and the electronic charge density, described by $v(\mathbf{r})$ and $j(\mathbf{r})$, respectively. The non-classical electron-electron interactions that give rise to $E_x^{HF}$ and $E_c$ in the Hartree-Fock approach are considered together in an effective potential $v_{xc}(\mathbf{r})$, the exchange-correlation potential. If the exchange operator $\hat{k}(\mathbf{x})$ in Eq. (5) is replaced by the term $ev_{xc}(\mathbf{r})$, one gets the so-called Kohn-Sham equations that can be solved iteratively, completely in analogy to the Hartree-Fock equations. In general, the resulting orbitals, the so-called Kohn-Sham orbitals, have a different meaning than the Hartree-Fock orbitals. Provided an exact expression is known for $v_{xc}(\mathbf{r})$, the squares of the Kohn-Sham orbitals add up precisely to the true ground-state density $\rho_0(\mathbf{r})$ and the energy of the highest occupied orbital equals the negative of the ionization energy (similar relations for the other orbital energies do not hold; there is no equivalent to Koopmans' theorem).

The total electronic energy in the Hartree-Fock scheme is given by $E^{HF}=E_{kin}^{HF}+E_{ne}^{HF}+E_{Coul}^{HF}-E_x^{HF}+E_c$, where the superscript denotes that Hartree-Fock orbitals are used for the evaluation of the individual contributions defined by Eqs. (7-11). In the Kohn-Sham scheme the total energy is given by $E^{KS}=E_{kin}^{KS}+E_{ne}^{KS}+E_{Coul}^{KS}-E_{xc}^{KS}$, where the exchange-correlation energy is defined by $E_{xc}^{KS}=\int v_{xc}(\mathbf{r})\rho(\mathbf{r})d\mathbf{r}$ and the superscript denotes that Kohn-Sham orbitals have been used in Eqs. (7-9). If the Hartree-Fock orbitals would be equal to the Kohn-Sham orbitals, the exchange-correlation energy $E_{xc}^{KS}$ could be interpreted as the sum of the exchange energy $E_x^{HF}$ and the correlation energy $E_c$ in the Hartree-Fock scheme. However, in real systems $E_{xc}^{KS}$ comprises also the differences $(E_{kin}^{HF}-E_{kin}^{KS})$, $(E_{ne}^{HF}-E_{ne}^{KS})$, and $(E_{Coul}^{HF}-E_{Coul}^{KS})$.

The Kohn-Sham scheme defines the functional $E_{xc}$ that represents exchange and correlation. A separation of $E_{xc}$ into exchange and correlation parts is not well defined. In practice, however, approximations to $E_{xc}$ are given as a sum of separate exchange and correlation functionals. If the exchange-correlation potential $v_{xc}(\mathbf{r})$ at a given position $\mathbf{r}$ depends only on the charge density $\rho(\mathbf{r})$ at that position [as in Eq. (12)], then $v_{xc}(\mathbf{r})$ is called *local*, and the functional $E_{xc}[\rho]$ belongs to the local density approximation (LDA). In retrospect the $X\alpha$ method can be regarded as the first LDA method and the success of this method is due to the fact that the exchange potential $v_x(\mathbf{r})$ in Eq. (12) contains a certain extent of correlation energy.

2.3 Generalized Gradient Approximation

The exchange operator $\hat{k}(\mathbf{r})$ defined in Eq. (6) is a non-local operator, because it depends on the value of the orbitals $\psi_i$ at all points in space. This fact has led to an improvement in the LDA by taking into account non-local effects. The first step in this direction is represented by exchange-correlation potentials of the form $v_{xc}[\rho(\mathbf{r}),\nabla\rho(\mathbf{r})]$, which include information about the gradient of the charge density. Surprisingly, first attempts to construct density functionals in this way did not lead to an improved accuracy. It turned out that additional restrictions (e.g. certain sum rules [3,6,7,19]) had to be enforced. The class of density functionals that are constructed in this way are known as generalized gradient approximations (GGA). These functionals have lead to major improvements and they are, together with the hybrid functionals (discussed in the next section), usually chosen if DFT methods are applied. Strictly speaking, these functionals are not non-local, because also $v_{xc}[\rho(\mathbf{r}),\nabla\rho(\mathbf{r})]$ depends on the properties of the charge density and its gradient at position $\mathbf{r}$ only. For this reason these functionals are sometimes called semilocal. A variety of different GGA functionals have been presented and applied in the last two decades (for an overview see Ref. [3]). Some popular GGA methods are Becke's exchange functional [20] together with the correlation functional of Lee, Yang, and Parr [21,22] (known as BLYP method), Perdew and Wang's exchange functional and their gradient-corrected correlation functional [23] (PW91 method), or Becke's exchange functional [20] together with Perdew's gradient-corrected correlation functional [24] (BP86).

2.4 Hybrid Functionals

The Hartree-Fock exchange energy $E_x^{HF}$, Eq. (10), can be calculated exactly, while in DFT only approximate exchange functionals are available. $E_x$ is usually about one order of magnitude larger than the correlation energy $E_c$, Eq. (11). Therefore, at first sight it seems to be promising to replace the approximate exchange functionals by $E_x^{HF}$ combined with an approximate correlation functional. Such a hybrid method indeed performs significantly better than the Hartree-Fock method but delivers poor results in comparison to GGA methods [3]. Obviously, by combining approximations for $E_x$ and $E_c$ a cancellation of errors occurs, which is not the case if the exact expression $E_x^{HF}$ is used. It should be noted also that for non-interacting Kohn-Sham orbitals the separation of exchange and correlation interaction is artificial. Despite this, combining the Hartree-Fock exchange with approximate density functionals has led to major improvements. Becke introduced a hybrid functional that includes 20 % Hartree-Fock exchange [25]. From this functional the currently most popular hybrid functional, B3LYP, was later derived [26] where the exchange correlation energy is defined as follows:

$$E_{xc}^{B3LYP} = (1-a)E_x^{LSD} + aE_x^{HF} + bE_x^{B88} + cE_c^{LYP} + (1-c)E_c^{LSD} , \qquad (13)$$

with the semiempirical parameters a=0.2, b=0.72, and c=0.81. With the B3LYP method, an average error of the atomization energy of less than 10 kJ mol$^{-1}$ has been obtained for the G2 test set [3]. A similar accuracy can be reached only with high-level wavefunction-based methods, which are computationally much more demanding. In its ability to predict the ground-state multiplicity of iron(II) SCO complexes, B3LYP usually fails. Instead, pure DFT methods predict the correct ground state but they exceed in favoring the LS state [27,28]. Reiher *et al.* ascribed this behaviour to the admixture of Hartree-Fock exchange and proposed a reparameterization of the B3LYP method by setting a=0.15 in Eq. (13) [29]. First applications of this new method, called B3LYP*, suggest that the reduced admixture of Hartree-Fock exchange leads to more accurate results for the energy splitting between states with different spin multiplicity. At the same time the B3LYP* method exhibits an accuracy of atomization and ionization energies that is comparable to B3LYP [30].

## 3 Applications

The application of DFT to SCO complexes is a fairly recent development. Only in the past few years have GGA and hybrid functionals emerged and speed and memory of modern computers increased such that DFT calculations on SCO complexes could be performed. First calculations have focussed on the frequency shift of the iron ligand bond stretching vibrations [31,32,33]. The calculation of intensities for IR, Raman, and nuclear inelastic scattering (NIS [2]) spectra enabled the interpretation of spectroscopic data on the basis of the calculated normal modes [34,35,36]. The reasonable agreement that has been obtained for calculated and measured frequencies has encouraged the calculation of vibrational entropy and energy differences [27,37,38]. One of the most difficult problems, the calculation of the difference of total electronic energy between HS and LS isomers, has been approached only very recently [28,38]. The reparameterization of the hybrid functional B3LYP brought the first methodological progress in this respect [29,30]. Still, the accuracy of the electronic energy difference is not sufficient to calculate transition temperatures exactly, but calculations for several iron(II) complexes with substituted pyrazolyl ligands have demonstrated that it is possible to predict qualitatively the effect of ligand substitution on the spin-transition temperature [27]. To do so, a simple model [39] has been used, that allows an explanation for gradual transitions, where the molar HS fraction $\gamma_{HS}(p,T)$ at given pressure $p$ changes smoothly over a large interval of temperature $T$:

$$\gamma_{HS}(p,T) = 1/[1 + \exp(\Delta G / k_B T)] \qquad (14)$$

Here $\Delta G$ denotes the difference (HS-LS) of the Gibbs free energy,

$$G = E_{el} + E_{vib} + pV - TS \qquad (15)$$

depending on the total electronic energy $E_{el}$, the vibrational energy $E_{vib}$, pressure $p$, volume $V$, temperature $T$, entropy $S$, and, implicitly, on $\gamma_{HS}$. This means that Eq. (14) has to be solved iteratively. The molar HS fraction $\gamma_{HS}$ is used for the determination of many properties of spin crossover materials, for instance the transition temperature $T_{1/2}$. It is defined by

$$\gamma_{HS}(p,T_{1/2}) = 1/2 \qquad (16)$$

The term $p\Delta V$ cannot be calculated in the molecular approximation that is usually applied. A typical value of $\Delta V \approx 7$ cm$^3$ mol$^{-1}$ [40] leads to a contribution to the free energy of $p\Delta V \approx 0.7$ J mol$^{-1}$ at ambient pressure and temperature. This is far less than the error margin of the other contributions to $\Delta G$, and the neglect of $p\Delta V$ is therefore justified.

## 3.1 Molecular Geometry and Mössbauer Parameters

For many SCO complexes, X-ray structures in either the LS or HS state or in both states are not available. In these cases, an optimization of the molecular geometry has to be the first step of a computational study. This means that the derivatives of the total energy with respect to the nuclear coordinates have to be calculated, and the coordinates have to be varied until the energy reaches a local minimum. There are several reasons why this procedure is also applied when X-ray structures are actually available. If calculated quantities for different complexes and spin states are compared, the differences might sensitively depend on the method that was used to determine the molecular geometry. This is clearly the case for differences of the total electronic energy (see section 3.4). In addition, most algorithms for frequency calculations demand a preceding geometry optimization. The comparison of measured and calculated geometries of SCO complexes is hampered by the fact that X-ray measurements are performed for solid samples whereas, to our knowledge, all calculations have been performed for free molecules. Some typical examples for measured and calculated iron ligand bond distances are given in Table 1. The iron ligand bond has been chosen because it belongs to the structural parameters of SCO complexes that are most difficult to calculate and because, to a first approximation, it can be regarded as the reaction coordinate of the spin crossover. Most of the calculated iron ligand bond distances given here are a few picometers smaller than the experimental values. Since this is true for both spin states, the deviations cancel out in part when the increase of the bond length ($\Delta$) upon spin crossover is calculated (Table 1).

## Table 1

Another test for the reliability of a computational study, besides the comparison of X-ray structure and calculated molecular geometry, is the investigation of Mössbauer parameters. Mössbauer spectroscopy is one of the key techniques for the investigation of iron(II) SCO complexes. Two basic parameters that can be obtained from Mössbauer spectra are the quadrupole splitting $\Delta E_Q$ and the isomer shift $\delta$. The latter reflects the total electronic charge density at the iron nucleus, while $\Delta E_Q$ provides information about the anisotropy of the electric field gradient at the iron center [41]. The quadrupole splitting can be qualitatively explained in the ligand-field model; it takes values of $\approx$ 1-4 mm s$^{-1}$ for an iron(II) HS complex and of $\approx$ 0 mm s$^{-1}$ for a LS complex. Electronic structure calculations are needed for a more accurate description. For iron(II) SCO complexes with substituted tris(pyrazolyl)methane ligands [42], measured and calculated data are compared in Table 2.

## Table 2

The parent complex of this series [27] consists of an iron(II) center and two tris(pyrazol-1-yl)methane ligands ([Fe(tpm)$_2$](PF$_6$)$_2$, compound **1a**, Figure 1). Additional complexes are obtained, if hydrogens of the pyrazole rings are substituted. Compounds **2a**, **2b**, and **2c** consist of an iron(II) center and two tris(3-methyl-pyrazol-1-yl)methane ligands [the superscripts denote the counterions PF$_6^-$ (a), ClO$_4^-$ (b), and BF$_4^-$ (c)]. Compounds **3a**, **4b**, and **5b** are derived from tris(4-methyl-pyrazol-1-yl)methane, tris(4-bromo-pyrazol-1-yl)methane, and tris(3,5-dimethyl-pyrazol-1-yl)methane, respectively.

## Figure 1

The calculated quadrupole splittings for the LS isomers are all close to zero (Table 2) as expected from the ligand-field model, while the measured values are about 0.3-0.4 mm s$^{-1}$ larger. One reason for this deviation is that the high symmetry of the free molecule ($D_{3d}$ according to the calculation) is reduced in the solid-state environment. The measured quadrupole splittings for the HS isomer are in the range from 3 to 4 mm s$^{-1}$, except for complex **1a**. For the latter complex, where the transition temperature is about 355 K, the quadrupole splitting of the HS state has been measured at room temperature only. The large deviation between the calculated and the measured $\Delta E_Q$ values for this complex is very likely due to a decrease of $\Delta E_Q$ with increasing temperature which is ascribed to the small splitting of the $t_{2g}$ orbital energies. The calculated quadrupole splittings of the SCO complexes **2-5** are in agreement with the experimental data. The small differences of $\Delta E_Q$ for complexes **2a**, **2b**, and **2c** illustrate the influence of the counterions, which is not included in the calculation for the free molecule.

In the case of temperature-dependent Mössbauer spectroscopy, the areas of the subspectra for the HS and LS isomers are used to determine the molar HS fraction $\gamma_{HS}$. For this purpose, the measured areas have to be corrected for the Lamb-Mössbauer factors $f_{LM}^{HS}$ and $f_{LM}^{LS}$, which can be quite different for HS and LS isomers. These factors can be determined experimentally by nuclear resonant forward scattering [2,35], or they can be calculated employing DFT methods. For molecular crystals the Lamb-Mössbauer factor can be approximated by a product of a lattice factor and a molecular factor [43]. For [Fe(tpa)(NCS)$_2$] a change of the molecular Lamb-Mössbauer factor from 0.92 (LS isomer at 34 K) to 0.75 (HS isomer at 107 K) has been calculated [35]. In order to calculate the complete Lamb-Mössbauer factor, calculations for a molecular crystal with periodic boundary conditions have to be performed.

## 3.2 Vibrational Modes and IR, Raman, and NIS Spectra

It is well known that the entropy change accompanying the crossover from the HS the LS state is positive. This entropy difference is due to the different spin state degeneracies and the changes of vibrational frequencies. The vibrational contribution to the entropy difference can be explained qualitatively by a simple model based on ligand-field theory [1]. Accompanying the transition from the LS to the HS state, two Fe 3d electrons are transferred from the $t_{2g}$ into the $e_g$ orbitals. Actually, these molecular orbitals are antibonding linear combinations of Fe 3d atomic orbitals and ligand atomic orbitals. Therefore, emptying the $\pi$-antibonding $t_{2g}$ molecular orbitals stabilizes the iron ligand bonds, whereas filling the more covalent $\sigma$-antibonding $e_g$ orbitals with electrons destabilizes these bonds. As a consequence, the frequency of the bond stretching vibrations is decreased. This frequency shift is generally difficult to observe by IR and Raman spectroscopy, since typical SCO complexes have a few hundred vibrational modes. It is very difficult to make a complete assignment of vibrational modes to the large number of observed lines. For the well-studied SCO complexes [Fe(phen)$_2$(NCS)$_2$] (phen = 1,10-phenanthroline) Takemoto et al. [44] reported a decrease of the iron ligand bond stretching frequency by a factor of two. Similar frequency shifts have been reported for several other SCO complexes [45]. This corresponds to a decrease of the force constant by a surprising factor of four. With the development of nuclear inelastic scattering (NIS) [2] a spectroscopic technique became available that focussed on the iron ligand bond stretching vibrations. At first NIS spectra were recorded for powder samples of the SCO complexes [Fe(tpa)(NCS)$_2$] (tpa = tris(2-pyridylmethyl)amine) [2,31,32,35] and [Fe(bpp)$_2$](BF$_4$)$_2$ [46] (bpp = bis(2,6-bis(pyrazol-3-yl)pyridine)), and later for [Fe(phen)$_2$(NCS)$_2$] [36]. Angular resolved measurements have been performed for a single-crystalline sample of [Fe(tptMetame)] [2,34] (tptMetame = 1,1,1-tris((N-(2-pyridylmethyl)-N-mythylamino)mythyl)ethane). Comparison of the measured NIS spectra with simulated spectra based on DFT frequency calculations for free molecules allowed the assignment of the observed peaks to normal modes, which have predominant iron ligand bond stretching character [32,34,36]. A perfectly octahedral metal complex should exhibit six metal-ligand bond stretching modes transforming according to the $A_{1g}$, $E_g$, and $T_{1u}$ irreducible representations of the octahedron group $O_h$. Since the *gerade* modes $A_{1g}$ and $E_g$ do not contribute to the mean-square-displacement of the metal center, these modes are not seen in the NIS spectra. In the spectrum of an ideally octahedral complex only one threefold degenerate $T_{1u}$ mode would be visible. Due to distortions of the molecular symmetry, the degeneracy of the $T_{1u}$ modes is lifted. If the complex does not possess inversion symmetry, the metal-ligand bond stretching modes that correspond to $A_{1g}$ and $E_g$ representations in octahedral symmetry can also become visible [2]. In Table 3 the frequencies of the bond stretching modes that have been observed in the NIS spectra of [Fe(tptMetame)], [Fe(tpa)(NCS)$_2$], and [Fe(phen)$_2$(NCS)$_2$] are given. The averaged increase of the bond stretching vibrations are 40 %, 47 %, and 31 %, respectively. This corresponds to an increase of the force constant by a factor of two. Comparison of the assignment made by Takemoto *et al.* [44] with the calculated normal coordinates suggests that one Fe-N-CS bending mode has been assigned as Fe-N bond stretching mode by Takemoto, leading thus to an increase of the Fe-N bond stretching frequency shift.

## Table 3

Apart from the assignment of the iron ligand bond stretching vibrations the comparison of the measured and simulated NIS spectra is of importance also as a quality test of the calculated normal modes. The NIS intensity depends on the eigenvalues and on the eigenvectors of the normal modes of vibrations. A similar test is provided by the comparison of measured and calculated IR and Raman intensities. For the complex [Fe(phen)$_2$(NCS)$_2$] the comparison of measured and calculated IR, Raman, and NIS intensities has been used to obtain a complete assignment of the 147 normal modes of vibration [36].

Detailed comparison of measured IR and Raman frequencies and calculated frequencies using a large basis set and different DFT methods have been performed by Reiher and coworkers [37,38]. They have demonstrated the importance of intermolecular interactions for an accurate calculation of the N-C stretching frequency of the isothiocyanates.

### 3.3 Entropy and Vibrational Energy Differences

In the previous section, it has been shown that the molecular vibrations of spin crossover complexes can be calculated with reasonable accuracy. Therefore, it is worthwhile to use the calculated vibrational modes to compute thermodynamic quantities such as the vibrational entropy and the vibrational energy. The temperature-dependent vibrational energy $E_{vib}(T)$ and vibrational entropy contribution $S_{vib}(T)$ of a molecule containing $M$ atoms can be calculated according to

$$E_{vib}(T) = k_B T \sum_{i=1}^{3M-6} x_i \coth(x_i) \quad (17)$$

and

$$TS_{vib}(T) = E_{vib}(T) - k_B T \sum_{i=1}^{3M-6} \ln[2\sinh(x_i)] \quad (18)$$

where $\omega_i$ is the angular frequency of vibrational mode $i$ and $x_i = \hbar\omega_i/2k_B T$. The vibrational energies have to be discussed on a purely theoretical basis, since there are no experimental data available for spin crossover complexes. The differences of vibrational energy influence the transition temperature according to Eq. (15), but the calculated results suggest that this influence is quite small. The situation is different for the entropy where experimental data are available for the difference between HS and LS states. Qualitative agreement can be obtained between the calculated entropy differences of free molecules and measured data for solid samples and for solutions (Table 4). Baranovic [47] has shown that the agreement can be significantly improved if solvation effects are taken into account in the calculation.

> **Table 4**

For the experimentally thoroughly investigated complex [Fe(phen)$_2$(NCS)$_2$] detailed calculations by Brehm *et al.* [37] and by Reiher [38] suggest that pure DFT methods like BP86 give the best agreement with experimental entropy differences. For the example of iron(II) complexes with tris(pyrazolyl)methane ligands, the extent to which the entropy difference is influenced by different substitutents at the ligands has been investigated [28]. The calculated differences are quite small (Figure 2) and it can be concluded that the observed changes in $T_{1/2}$ are due mostly to changes of the total electronic energy differences $\Delta E_{el}$.

> **Figure 2**

## 3.4 Electronic Energy Differences

The total electronic energy difference between the HS and the LS isomer, $\Delta E_{el} = E_{el}^{(HS)} - E_{el}^{(LS)}$, is one of the most basic and yet most difficult to calculate quantities associated with SCO. Taking the entropy difference and transition temperature $T_{1/2}$ from experiment, the total energy difference can be roughly estimated by the relation

$$\Delta E_{el} \approx T_{1/2} \Delta S(T_{1/2}) \ , \qquad (19)$$

leading to values in the range of 1 to 30 kJ mol$^{-1}$. The calculated electronic energy of a typical SCO complex is of the order of some 10$^6$ kJ mol$^{-1}$. The energy difference is therefore, more than five orders of magnitude smaller than the absolute energy values. Since the total electronic energy of a transition metal complex cannot be calculated with an accuracy of some kJ mol$^{-1}$ (corresponding to a relative error of about 1 ppm), a successful calculation of $\Delta E_{el}$ depends on an extensive cancellation of errors. In other words, it is important that errors, which cannot be avoided, affect the calculated energy of both isomers in the same way. Due to the small energy difference between LS and HS isomers, calculations for SCO complexes are extremely sensitive to a bias towards a particular spin multiplicity. This makes SCO complexes ideal objects for testing electronic structure methods that should be able to predict the true ground state multiplicity of transition metal complexes. For transition metal complexes containing up to one hundred atoms, there are no high-level ab initio calculations available which serve as a benchmark, and therefore the results of any method can be judged only in the light of experimental results.

Concerning the spin multiplicity of iron(II) complexes, there are two classical examples, $[Fe(H_2O)_6]^{2+}$ and $[Fe(CN)_6]^{4-}$, which exemplify the two extreme cases, namely the diamagnetic, LS $^1A_{1g}(t_{2g}^6)$ ground state and the paramagnetic, HS $^5T_{2g}(t_{2g}^4e_g^2)$ ground state [1]. For both complexes, the geometry of the free molecule has been optimized for the HS and for the LS state, using the Hartree-Fock method and various DFT methods. The calculated energy difference between the HS and the LS state (Table 5) shows that any of the applied methods yields the correct ground state for these two complexes. However, large differences are observed in the results of the individual methods. For the SCO complex $[Fe(tpm)_2]^{2+}$ (complex **1**), for example, the differences between the computational methods become even more obvious.

Unfortunately, most applicable methods have a bias towards either the LS or the HS state. This is most obvious in the case of the Hartree-Fock method, which systematically favors higher spin multiplicities. The reason is that in this method the correlation between electrons with the same spin projection (Fermi correlation) is taken into account by the exchange interaction. Correlation between electrons with different spin projections (Coulomb correlation) is instead completely neglected in the Hartree-Fock scheme. For example, the four unpaired 3d electrons in an iron(II) HS complex avoid each other to a large extent due to the orthogonality of the 3d orbitals. In the LS complex, the three lowest 3d orbitals are doubly occupied and due to the monodeterminantal form of the wavefunction the spin-up and spin-down electrons occupying the same 3d orbital cannot avoid each other to the same extent. As a result, in the LS case the Coulomb repulsion between the electrons is too large.

The situation is better for DFT methods, which include some correlation effects, and in general, DFT methods predict the correct LS ground state for SCO complexes. However, pure DFT methods seem to favor the LS state leading to an energy difference $\Delta E_{el}$ which is too large [27,29]. Currently, the most accurate functionals for the calculations are hybrid functionals, namely the B3LYP method. Although the B3LYP method often predicts the wrong ground state multiplicity [27,28,29,30], the deviation between the calculated $\Delta E_{el}$ and the true energy difference seems to be smaller than in the case of pure density functionals. The most accurate functional for this purpose is the reparameterized hybrid functional B3LYP* [29,30] where the admixture of the Hartree-Fock exchange has been reduced in comparison to the B3LYP method.

For the calculation of vibrational frequencies, it has been found out that the accuracy of the calculated values can be improved if experimental geometries are used instead of optimized ones [48]. It is an appealing idea to use the same procedure for the calculation of electronic energy differences. The drawbacks of this procedure are illustrated with the SCO complex [Fe(tpen)](ClO$_4$)$_2$. Chen et al. [49] obtained for the HS and the LS isomer of this complex total energies of -6789.985 and -6790.071 MJ mol$^{-1}$, respectively, using the B3LYP/3-21G method and an X-ray structure (site A) at 293 K [49]. The calculations were performed for free molecules. Calculations using the same DFT method and using the X-ray structure (site A) at 298 K given in [50], yielded total energies of -6790.013 and -6790.048 MJ mol$^{-1}$ [27]. Both calculations are in general agreement, but the resulting total energy differences (85 kJ mol$^{-1}$ [49] and 35 kJ mol$^{-1}$ [27]) deviate significantly. Since both calculations have been performed with identical methods implemented in the same program package, Gaussian98 [51], this deviation is probably due to very small differences in the X-ray structures given in Refs. [50] and [49]. After optimization of the geometry of the free cation [Fe(tpen)]$^{2+}$ using B3LYP/3-21G, total energies of -6790.168 MJ mol$^{-1}$ and -6790.158 MJ mol$^{-1}$ were obtained for the HS and the LS isomer, respectively [27]. The resulting energy difference $\Delta E_{el}$=-10 kJ mol$^{-1}$ suggests a HS ground state which is wrong. However, considering the error margin of these calculations (especially in view of the very small basis set) the results of all three calculations, using the X-ray structures in [50] and [49] and the optimized geometries in [27], are in agreement. The large deviations for $\Delta E_{el}$ that are traced back to the errors of the X-ray structures do not support the assumption that the accuracy of $\Delta E_{el}$ could be improved significantly by the use of X-ray structures.

The comparison of calculated electronic energy differences with experimental results is hampered by the fact that the large majority of experimental data has been gained for solid state samples. The electronic energy difference that has been derived from experimental data is therefore influenced by intermolecular interactions which are not present in calculations for free molecules. From the observed shifts of the transition temperatures when replacing counterions [28,52,53] it can be concluded that the influence of the intermolecular interactions is comparable in size with the error of the calculated $\Delta E_{el}$ when using for instance B3LYP*. The next step to increase the accuracy of $\Delta E_{el}$ should therefore be calculations with periodic boundary conditions.

**Table 5**

## 3.5 Substituent Effects on the Transition Temperature

Due to the insufficient accuracy of the calculated total energy difference $\Delta E_{el}$, it is not possible to predict the transition temperature of a SCO complex directly from theory. The situation can be different, if the shift of transition temperatures is of interest, e.g. when going from one complex to another very similar one. While the calculated absolute temperatures for these complexes are meaningless, the calculated temperature difference can be a reasonable value. This can happen, when the errors of the calculated absolute temperature cancel to a large extent. Ideal objects for such a computational strategy are, for instance, iron(II) complexes with tris(pyrazolyl)methane ligands [42]. Some of these complexes have been studied by Mössbauer spectroscopy and magnetic susceptibility measurements, and from the observed spectra the molar high-spin fraction and the transition temperature have been extracted [28]. All substituents, except for bromine, lead to a decrease of the transition temperature. Density functional calculations have been carried out to compare the experimentally observed shifts of the transition temperature with those derived from theory. These calculations have been carried out for isolated complexes *in vacuo*. This approximation neglects interactions between neighboring complexes and interactions between the complexes and their counterions. Both interactions are known to considerably influence $T_{1/2}$. Calculations that do not regard these interactions can therefore at best be in qualitative agreement with the experiment. Nevertheless, such calculations for isolated complexes *in vacuo* may reveal information about the molecular contribution to substituent-induced shifts of $T_{1/2}$. This information can hardly be gained experimentally since any experiment with a solid sample will only reflect the combined influence of intra- and intermolecular interactions.

For the complexes **1a**, **2a**, **2b**, **2c**, **3a**, **4b**, and **5b**, which were introduced in section 3.1, the transition temperatures have been determined by Mössbauer spectroscopy and magnetic susceptibility measurements [28]. For the compounds **2a** and **5b** the transition temperature determined by magnetic susceptibility measurements is close to 0 K and it might be that these complexes undergo only a partial or no spin transition. Except for compound **4b**, which has roughly the same transition temperature as the parent complex **1a** (≈ 355 K), all other compounds exhibit a lower transition temperature. Substituting a hydrogen atom of the pyra-zol ring by a methyl group, as in compounds **2a**, **2b**, **2c**, and **3a**, decreases $T_{1/2}$ in comparison to the transition temperature of **1a**. The decrease is even greater if two hydrogens are substituted, as is the case for compound **5b**. Comparison of compounds **2a**, **2b**, and **2c** reveals a significant influence of the counteranion on $T_{1/2}$.

From the DFT calculations for complexes **1** to **5** the difference of electronic energy $\Delta E_{el}$, of vibrational energy $\Delta E_{vib}$, and of entropy $\Delta S$ can be retrieved. It has been assumed here, that the lowest electronic excitation energy is large compared to $k_B T$, and hence $\Delta E_{el}$ is regarded to be constant in the temperature range of interest. All calculated terms of the free energy depend significantly on the chosen method and basis set.

The temperature-dependent free energy difference $\Delta G(T)$ is calculated by summing up the terms given in Eq. (15) except for the volume change, which is neglected. Ideally, the points of intersection of the curves $\Delta G(T)$ with the abscissa should yield the transition temperatures $T_{1/2}$, according to its definition in Eq. (14). However, it is obvious that the calculated curves are shifted due to errors of $\Delta E_{el}$, and this prevents the determination of absolute values for $T_{1/2}$. A rough estimate for the difference $\Delta T_{1/2}$ of transition temperatures when comparing two complexes **a** and **b** can be obtained by the expression

$$\Delta T_{1/2} \approx \left(\Delta E_{el}^b - \Delta E_{el}^a\right) / \Delta S\left(T_{1/2}^a\right) \qquad (20)$$

where $T_{1/2}^{\mathbf{a}}$ must be known from experiment and $\Delta S^{\mathbf{a}}(T) \approx \Delta S^{\mathbf{b}}(T)$ is assumed. Comparison of the shift of the transition temperature estimated in this way with experimental values yields agreement for the direction of the shift and for the order of magnitude of the shift.

It has been shown for the example of bis-tripodal chelates of iron(II) with tris(pyrazolyl)methane ligands that $T_{1/2}$ can be influenced significantly, if methyl groups substitute hydrogen atoms of the pyrazole rings. The calculations presented here suggest that it should be possible with currently available density functional methods to predict the direction and the order of magnitude of a shift of the transition temperature.

It is interesting to note that the calculated shifts of $\Delta E_{el}$, accompanying the change from the unsubstituted complex **1** to the substituted complexes, seem to be of similar order of magnitude for the HF methods and for the DFT methods (Table 6, values in brackets). It seems that, when comparing similar complexes, a large part of the error of $\Delta E_{el}$ cancels out.

**Table 6**

## 4 Summary

Density functional methods have been proven to provide fairly accurate results for many properties of SCO complexes. Even with modest basis sets, the geometrical structure can be described qualitatively. Using large basis sets, an accuracy of a few picometers can be obtained for bond lengths. Good agreement has been reached also for Mössbauer parameters, vibrational frequencies, and entropy differences. Due to the reparameterized hybrid functional B3LYP* an accuracy of about 10 kJ mol$^{-1}$ can be reached for the total electronic energy difference. This is unfortunately still not sufficient for direct calculations of the transition temperature. However, DFT calculations can provide reasonable estimates for transition temperature differences, if complexes are investigated that differ only by small modifications of the ligands.

It is expected that calculations for solid samples using periodic boundary conditions will bring further progress in the field of electronic structure calculations for SCO complexes.

*Acknowledgements*: This work was supported by the TMR project TOSS (ERB-EMRX-CT98-0199) and by the Deutsche Forschungsgemeinschaft within the priority programme „Molecular Magnetism". We thank L. Duelund, A. Zimmermann, and H. Toftlund for the fruitful cooperation.

Figure 1. Molecular structure of the HS isomer of complex 1 calculated with B3LYP/6-311G. Hydrogens have been omitted for clarity.

Figure 2. Temperature dependent vibrational entropy difference for complexes 1, 2, and 4 calculated with BLYP//LANL2DZ (taken from Ref. [28]).

Table 1. Measured and calculated iron ligand bond distances in Å.

| Complex | Method | Fe-$N_{NCS}$ | | | Fe-$N_{lig}$ | | |
|---|---|---|---|---|---|---|---|
| | | LS | HS | Δ | LS | HS | Δ |
| [Fe(phen)$_2$(NCS)$_2$] | X-ray [54] | 1.96 | 2.06 | 0.10 | 2.00 | 2.20 | 0.2 |
| | DFT | 1.90 | 1.98 | 0.08 | 1.93 | 2.17 | 0.24 |
| [Fe(tpa)(NCS)$_2$] (β) | X-ray [55] | 1.95 | 2.08 | 0.13 | 1.96 | 2.21 | 0.25 |
| | EXAFS [35] | 1.95 | 2.05 | 0.10 | 1.95 | 2.19 | 0.24 |
| | DFT | 1.91 | 2.00 | 0.09 | 1.93 | 2.23 | 0.30 |
| [Fe(bptn)(NCS)$_2$][a] | X-ray [56] | | 2.10 | | | 2.20 | |
| | DFT | 1.91 | 2.03 | 0.12 | 1.94 | 2.18 | 0.24 |
| [Fe(tptMetame)] | X-ray [57] | | | | 2.04 | 2.24 | 0.20 |
| | DFT | | | | 2.09 | 2.25 | 0.16 |

[a] bptn = N,N'-bis(2-pyridylmethyl)-1,3-propanediamine

Table 2. Measured quadrupole splitting $\Delta E_Q$ and isomer shift $\delta$ in mm s$^{-1}$ (values calculated with B3LYP/6-311G in parentheses)[a]

| Complex | $\Delta E_Q$ | | $\delta$ | |
|---|---|---|---|---|
| | LS | HS | LS | HS |
| **1a** | 0.30 (0.10) | 2.20 (3.81) | 0.47 | 0.85 |
| **2a** | 0.39 (0.01) | 3.96 (3.75) | 0.53 | 1.13 |
| **2b** | 0.43 (0.01) | 3.71 (3.75) | 0.53 | 1.24 |
| **2c** | 0.36 (0.01) | 3.99 (3.75) | 0.53 | 1.13 |
| **3a** | 0.32 (0.09) | 3.55 (—) | 0.51 | 0.97 |
| **4b** | 0.35 (0.12) | 3.12 (3.34) | 0.48 | 1.41 |
| **5b** | — (—) | 3.99 (—) | — | 1.15 |

[a] all values taken from Ref. [28]

Table 3. Iron ligand bond stretching frequencies in cm$^{-1}$ observed by NIS.

| Complex | Ref. | HS | LS |
|---|---|---|---|
| [Fe(tptMetame)] | [34] | 226, 266 | 323, 331, 355, 371 |
| [Fe(tpa)(NCS)$_2$] | [32] | 242, 282 | 347, 379, 427 |
| [Fe(phen)$_2$(NCS)$_2$] | [36] | 229, 331 | 359, 366, 377 |

Table 4. Calculated and measured entropy differences in J mol$^{-1}$ K$^{-1}$

| Complex | $\Delta S$ | | $T_{1/2}$ |
|---|---|---|---|
| | Exp. | DFT | |
| [Fe(phen)$_2$(NCS)$_2$] | 49 [58] | 60 | 176 |
| [Fe(tacn)$_2$]$^{2+}$ in D$_2$O [a] | 61(3) [59] | 55 | 344 |
| [Fe(tacn)$_2$]$^{2+}$ in acetone [a] | 73(37) [59] | 55 | 328 |

[a] tacn = 1,4,7-triazacyclononane

Table 5. Energy difference $\Delta E$ (in kJ mol$^{-1}$) between HS and LS states.

| | [Fe(H$_2$O)$_5$]$^{2+}$ | [Fe(tpm)$_2$]$^{2+}$ | [Fe(CN)$_6$]$^{4-}$ |
|---|---|---|---|
| HF | | -300.158[a] | -1815.57545854 |
| MP2 | | -137.178[b] | |
| HFS | | | |
| BLYP | | 81.721[a] | 139.92 |
| PW91 | | 103.552[a] | -1820.27187594 |
| B3LYP | -562.409 (D$_{2h}$) | -7.302[a] | 0.607 |
| | -238.282 (C$_{2h}$) | | |
| B3LYP* | | 21.473 | 49.418 |
| Experiment[c] | << 0 | ≈26 | >> 0 |

[a] taken from Ref. [28]

[b] using the geometry optimized with HF and employing the LANL2DZ basis set

[c] estimation based on experimental $T_{1/2}$ and calculated $\Delta S$ as explained in the text

Table 6. $\Delta E_{el}$ in kJ mol$^{-1}$ calculated with the 6-311G basis set and with the Hartree-Fock and different DFT methods (values in brackets give the difference to complex **1**)

| Complex | HF | B3LYP | B3LYP* | BLYP | PW91 |
|---|---|---|---|---|---|
| 1 | -300.16[a] | -7.30[a] | 21.47 | 81.72[a] | 110.77[a] |
| 2 |  | -39.51[a] | -12.71 | 39.12[a] | 71.33 |
|  |  | (32.21) | (34.18) | (42.60) | (39.44) |
| 3 |  | -6.13 | 22.27 | 83.40 | 111.98 |
|  |  | (-1.17) | (-0.80) | (-1.68) | (-1.21) |
| 4 | -300.67 | -7.72 | 19.83 | 79.12 | 107.64 |
|  | (0.51) | (0.42) | (1.64) | (2.60) | (3.13) |

[a] taken from Ref. [28]

Fig. 1

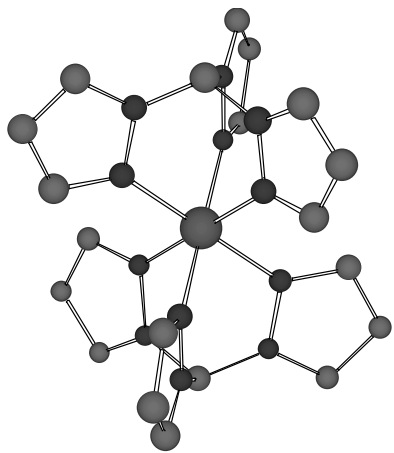

Fig. 2

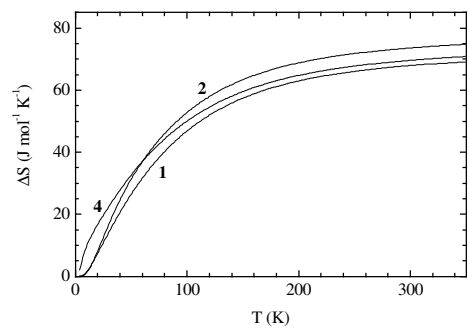